# Driver Heterogeneity in Willingness to Give Control to Conditional Automation

Muhammad Sajjad Ansar[1], Nael Alsaleh[2], and Bilal Farooq[1]
[1]Laboratory of Innovations in Transportation (LiTrans), Toronto Metropolitan University, ON, M5B 2K3, Canada
[2] Department of Civil & Mineral Engineering, University of Toronto, ON, M5S 1A4, Canada

**The driver's willingness to give (WTG) control in conditionally automated driving is assessed in a virtual reality based driving-rig, through their choice to give away driving control and through the extent to which automated driving is adopted in a mixed-traffic environment. Within- and across-class unobserved heterogeneity and locus of control variations are taken into account. The choice of giving away control is modelled using the mixed logit (MIXL) and mixed latent class (LCML) model. The significant latent segments of the locus of control are developed into *internalizers* and *externalizers* by the latent class model (LCM) based on the taste heterogeneity identified from the MIXL model. Results suggest that drivers choose to 'giveAway' control of the vehicle when greater concentration/ attentiveness is required (e.g., in the nighttime) or when they are interested in performing a non-driving-related task (NDRT). In addition, it is observed that internalizers demonstrate more heterogeneity compared to externalizers in terms of WTG.**

## I. INTRODUCTION

In SAE Level-3 vehicles [1], a human driver has a choice to give away the control of the dynamic driving tasks (DDT) to the automated driving system. Upon relinquishing the driving control by a human driver, conditionally automated driving (CAD) can take over the driving tasks in the prescribed operational design domains (ODD). However, ensuring the individual's willingness to give control (WTG) in a futuristic traffic environment can not be guaranteed solely based on the willingness to pay for L3 vehicles or having the intention to use them. Furthermore, the decision to transfer the control is entirely voluntary, and it thus entails heterogeneity in usage behaviour.

One of the personality traits used in explaining how drivers adopt automated driving is the locus of control. For example, Rudin et al. [2] estimated the impact of a driver's locus of control on the behavioural adaptation of an automated driving feature of lane departure warning. In general, individuals with an internal locus of control, i.e., *Internalizers*, believe that they can control events, whereas, *Externalizers*, those with an external locus of control, believe external factors control events in their lives [3]. In particular, it is hypothesized in the literature that individuals categorized as internalizers are less inclined to adopt automated driving compared to externalizers. For instance, a study by Gabrhel et al. [4] explored the assumption that the internalizers would exhibit lower trust toward automated driving as they rely on and have confidence in their own driving abilities. However, the validity of this assumption could not be established, as the original two-factor externality-internality structure of the locus of control did not demonstrate a satisfactory fit with their available data. Moreover, the impact of locus of control on driving behaviour cannot be disregarded, as it can also be accompanied by the presence of heterogeneity effects.

Econometric discrete choice models offer great potential in modelling behavioural response by providing a framework to understand and predict how individuals make choices among various options [5]. These models can investigate the factors influencing a driver's WTG, measure their effects on choices, and evaluate driver preferences. To address the unobserved heterogeneity, Mixed logit (MIXL) and latent class (LC) models offer two different ways. LC model categorizes individuals into discrete classes of subpopulations according to their observed characteristics and behaviour, whereas the MIXL model uses random effects that vary between individuals to capture individual-level variations. LC model incorporates the distinct subpopulations within a population and enables the modelling of conditional dependencies between observed variables within each latent class. LC probabilistically divides the population into certain latent classes by defining class-membership functions and class-specific utility functions. The combined analysis of LC and MIXL enhances a deeper understanding of behavioural heterogeneity, diverse preferences, and decision-making patterns [6].

The individual's WTG to automated driving can be influenced by their previous experiences, sociodemographic characteristics, perceptions, and attitudes [7]. Moreover, driving settings involving different traffic environments, weather, and lighting conditions could also considerably impact WTG [8]. Different kinematic parameters represent various driving styles that could also potentially contribute to the behavioural modelling of an individual's WTG [9]. Given that, the primary objective of the present study is to estimate the driver's WTG under different settings in a mixed-traffic environment. We employed an exploratory study method following the prevalent approach for studying combined conditions, typically involving short driving scenes with experiments lasting under five minutes [10]. In addition, investigating taste heterogeneity among individuals, exploring the subgroups among the population, and addressing differences in response probabilities among individuals in subgroups, are sub-objectives of this study.

In this study, an individual's WTG is analyzed by considering two dependent variables (DVs): the binary decision of giving or not giving away control to automated driving and the extent of automated driving adoption assessed by the proportion of time participants spend using automated driving within an L3 vehicle trip. The binary choice is estimated through the MIXL and mixed latent class (LCML) model, having the binary logit as a base mode. The second DV is the levels of willingness based on the duration of automated driving in a trip in a mixed-traffic environment. Finally, an ordinal logit (OL) model is estimated to analyze the low, medium, and high levels of automated driving adoption. The current study investigates the following key research questions:

1) What factors are associated with the choice of giving away control and the extent to which drivers adopt automated driving in a mixed-traffic environment?
2) How can the individual and subpopulation level heterogeneity be effectively modelled?
3) What are the distinct behavioural patterns and characteristics exhibited by the identified subpopulation, and how do these patterns influence the choice of giving away control?

The subsequent sections of the paper are organized as follows: Section II provides a background on the exploratory simulator studies and the approaches used in the literature for modelling the heterogeneity. Section III discusses the experimental setup and presents descriptive statistics of the data. Section IV outlines the methodology employed in this study. Section V provides the implementation details. The model results and their significance are presented in Section VI. Finally, Section VII concludes the paper with a discussion and summary of the findings.

## II. BACKGROUND

This section highlights the importance of exploratory simulator studies and provides an overview of different virtual immersive reality setups utilized in the underlying research domain. Additionally, it provides an overview of hybrid models for modelling behaviour heterogeneity. Finally, concluding remarks are provided at the end of this section.

### A. Exploratory Simulator Studies

The willingness towards automated driving is frequently examined based on individuals' stated intentions, but the evaluation of its actual usage in real-world scenarios is less commonly explored. For instance, O'Hern and Louis [11] evaluated the participant's readiness and intentions to use conditionally automated vehicles through an online survey. Self-reported surveys, questionnaires, and hypothetical scenarios often encounter issues such as hypothetical bias and a lack of realism, as they might not adequately represent behaviours in the real world and can create biases. Moreover, knowledge and experience can alter the trust and preferences regarding the willingness to use automated vehicles [12]. So, to improve the quality of outcomes, some studies have taken a proactive approach by introducing AVs to the participants before conducting surveys related to their adoption. For example, Charness et al. [13] first familiarized the participants with AVs by visual depiction and written description. Then, they gathered responses about their attitudes toward adopting AV technology and their willingness to relinquish driving control. Similarly, Ayoub et al. [14] investigated the self-reported driver's trust in CAD by showing them successful and unsuccessful takeover driving scenes. However, exploratory simulator studies are a step forward in reducing hypothetical bias and increasing spatial knowledge realism. These studies provide safe and controlled settings for examining how different driving scenarios, and automated driving features affect driver behaviour [15]. For instance, implementing a setup such as VIRE (Virtual Immersive Reality Environment) [16] enables researchers to build controlled experimental settings for analyzing driving behaviour towards automated vehicles in futuristic traffic environments.

### B. Integration of Digital Twin and VR for Immersive Driving

In recent years, driving behaviour analysis has dramatically improved by integrating virtual reality (VR) and digital twins (DT). This innovative approach combines DT-based automated vehicles with a VR-generated mixed traffic environment to develop controlled experimental settings. Using VR, real-world driving experiences can be created that offer a realistic and immersive environment. More trustworthy data can be acquired since drivers can engage with futuristic traffic environments and experience various driving settings. In contrast, DT refers to a virtual mapping technology that can generate complete or partial construction of the vehicle's virtual model. In the domain of automated driving vehicles, the novel architecture of DT technology allows an understanding of drivers' interactions with precisely mirrored automated features and decision-making processes in a more objective and thorough manner [17].

Automated DT systems utilized in the literature range from low to high fidelity, characterized by their capacity to simulate real-world driving conditions and deliver realistic motion and sensory feedback to drivers, with consequential effects on driving behaviour and performance [18]. High-fidelity DT system could only be obtained by motion-based simulators with multi-degree-of-freedom (DoF) characteristics, 360° visual immersion, and long-lasting accelerations for all driving manoeuvres, for instance, in [19]. However, efforts are deployed in different ways to create immersive driving scenes depending on the available resources, the specific context and the requirements of the research task. For example, fixed-based driving simulators, which are based on stationary platforms and lack the full motion capabilities of high-fidelity simulators, continue to be widely used despite their limited physical motion and a lack of haptic feedback. Pan et al. [20] used a fixed-based driving simulator with a 27-inch display screen to analyze the impact of non-driving related tasks on driver sleepiness and takeover performances in prolonged conditional automated driving. The fixed-based driving simulator setup used by Ross et al. [21] had a better immersion with three 4K display screens that analyzed the impact of decorated traffic enforcement cameras on safety. Moreover, Oh et al. [22] significantly improved the fidelity of the fixed-based driving simulator by incorporating a head-mounted display (HMD) to enhance visual immersion and by integrating microscopic traffic flow parameters into the virtual environment to improve realistic driving conditions. The underlying VIRE as a DT system is a viable alternative to high-fidelity simulators [23]. It outperforms the above-listed fixed-based driving simulators as it entails four major systems, including a scenario development system, multi-model traffic micro-simulation system, virtual environment projection system, and response tracking system.

### C. Modelling Behaviour Heterogeneity

The simultaneous estimation of hybrid models for modelling behaviour heterogeneity is primarily used for two purposes in the literature. Some studies compare hybrid models' outcomes to find the preferred model. For example, Cerwick et al. [24] employed simultaneous estimation of MIXL and LC models to identify the optimal one to address the unobserved heterogeneity. The mixing effects of contributing factors were estimated only in the MIXL model. They aimed to determine the preferred model for developing the crash severity model. Similarly, Zhang et al. [19] estimated the hybrid models with and without the non-decreasing function of driving time for drowsy driving analysis. They compared the drowsiness detection accuracy of mixed-effect ordered logit models to find the best detection model.

In contrast, few studies integrate the outcomes of hybrid models, which can enhance predictive capabilities and improve model performance. For example, Alsaleh et al. [5] simultaneously estimated the hybrid models and combined the latent class (LC) analysis with integrated choice and latent variable (ICLV) into the latent class integrated choice and latent variable (LC-ICLV) model. The objective was to analyze the preference of public transit users towards On-demand transit across the population, taking into account human subjectivity and the behavioural heterogeneity across the population's latent segments. Similarly, Han and Timmermans [6] have addressed the inherent heterogeneity by simultaneously estimating hybrid models. Their LC analysis identified the sub-classes and the MIXL model explored the unobserved heterogeneity within each latent class. Their integrated analysis of LC and MIXL (referring to "LCML" in this paper) addressed the differences in response probabilities among individuals of subgroups besides uncovering the segmentation in the population.

### D. Concluding Remarks

While some studies have explored factors like trust in automation, willingness to pay for L3 vehicles, and intention to use automated driving, there is still a lack of thorough research into the underlying aspects that affect the individual's willingness to use automated driving control at different levels. To the best of the authors' knowledge, systematic analysis of the individual's WTG has yet to be addressed in the literature. Moreover, the literature on hybrid choice analysis suggests that incorporating individual behavioural variables and accounting for behavioural heterogeneity across population segments in modelling techniques can enhance the explanatory power of traditional models. To that end, the WTG analysis in controlled environment settings based on locus of control segmentation could offer valuable insights to policymakers in better identifying the target market segments and suggesting segment-specific recommendations. The present analysis would potentially facilitate a comprehensive assessment of trust in and acceptance of automated driving mode. It would assist in enhancing the safety and efficiency of mixed-traffic environments.

## III. Data

In this section, the discussion begins with exploring the experimental design and the employed control variables. It then explores the data collection and presents descriptive statistics.

### A. Controlled Variables and Repetition of Experiments

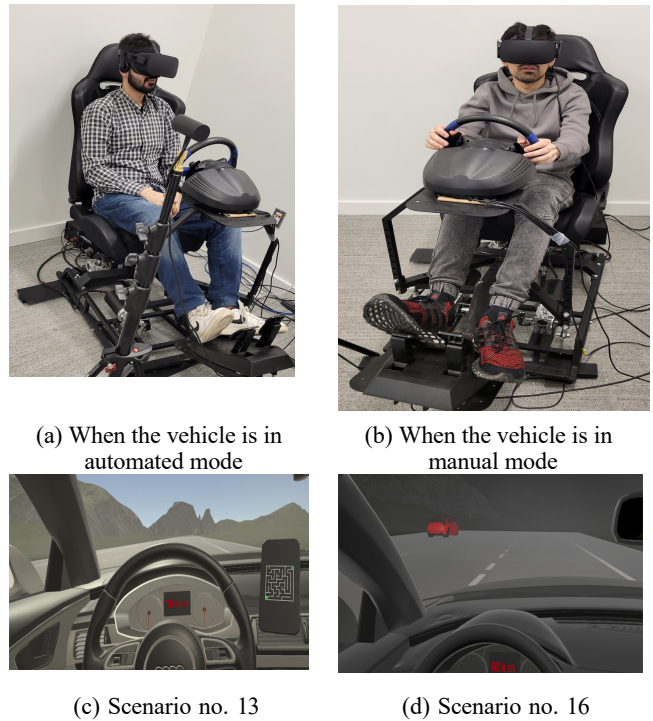

(a) When the vehicle is in automated mode

(b) When the vehicle is in manual mode

(c) Scenario no. 13

(d) Scenario no. 16

Fig. 1: Driving experience in VIRE.

To examine the conditions under which the drivers are more inclined to share driving control with automation, controlled laboratory experiments were conducted using VIRE [16]. Sixteen scenarios, as outlined in Table I, were employed for this purpose. The experiments focused on investigating the voluntary switching of driving control. The term "giveAway" denoted scenarios where drivers shared control with automation (as depicted in Figure 1a), while "no_giveAway" represented driving scenarios without control sharing (as depicted in Figure 1b).

Each scenario consists of four controlled variables, with two levels assigned to each variable. Figure 1c and Figure 1d present two example scenarios to depict how the control variables look like in the underlying scenarios. On average, each scenario is replicated nearly ten times, and each replication lasts approximately 5 minutes. Interestingly, drivers opted to share control with automation in 75 percent of the experiments (129 out of 172). This indicates that one out of every four experiments was solely run in manual mode, without any control sharing (i.e., no_giveAway). Scenarios three, seven, and fifteen stand out as having the highest frequency of "giveAway" responses, indicating that both nighttime driving and multi-tasking exert a notable positive impact on WTG. On the other hand, scenarios fourteen and sixteen demonstrate the highest occurrence of "no_giveAway" responses, suggesting that rainy weather combined with light congestion is less favourable for promoting control sharing.

### B. Participants

The campaign of experiments was conducted in two waves. In the initial wave [25], the data on both taking back control and giving away control were collected from a total of forty-seven (47) participants. Each participant was assigned four (4) randomly selected scenarios that included both types of control situations. It contributed around ninety (90) "giving away" observations. In order to expand the "giving away" responses as per the current study requirements, an additional twenty-one (21) participants were recruited in the second wave, resulting in eighty-two (82) additional "giving away" observations. Here, the participants were assigned only "giving away" scenarios, following the same protocols as adopted in the previous wave. The objective was to ensure a minimum of ten repetitions for each scenario in the "giving away" experiments. Overall, data from a diverse group of sixty-eight (68) participants with various sociodemographic characteristics resulted in a total of 172 observations. Apart from the experiments exclusively focused on "giving away" scenarios, the overall experimental setup, including information and learning sessions, remained consistent with the approach used in our recent study [23].

TABLE I: Controlled variables and repetition of experiments in the given scenarios

| Scenario | Weather | | Lighting | | Multi-tasking | | Traffic Congestion | | Observations (total = 172) | | |
|---|---|---|---|---|---|---|---|---|---|---|---|
| | Clear Sky | Rainy | Day | Night | Yes | No | Heavy | Light | giveAway | no_giveAway | total |
| 1 | x | | x | | x | | x | | 10 | 4 | 14 |
| 2 | x | | x | | | x | x | | 8 | 3 | 11 |
| 3 | x | | | x | x | | x | | 11 | 0 | 11 |
| 4 | x | | | x | | x | x | | 7 | 4 | 11 |
| 5 | x | | x | | x | | | x | 7 | 3 | 10 |
| 6 | x | | x | | | x | | x | 9 | 1 | 10 |
| 7 | x | | | x | x | | | x | 8 | 0 | 8 |
| 8 | x | | | x | | x | | x | 6 | 3 | 9 |
| 9 | | x | x | | x | | x | | 7 | 4 | 11 |
| 10 | | x | x | | | x | x | | 6 | 4 | 10 |
| 11 | | x | | x | x | | x | | 9 | 2 | 11 |
| 12 | | x | | x | | x | x | | 8 | 3 | 11 |
| 13 | | x | x | | x | | | x | 9 | 1 | 10 |
| 14 | | x | x | | | x | | x | 5 | 5 | 10 |
| 15 | | x | | x | x | | | x | 12 | 1 | 13 |
| 16 | | x | | x | | x | | x | 7 | 5 | 12 |

TABLE II: Description of variables with giveAway percent

| Variables | Intervals | Variable Definition | giveAway (%) | Obsevations (N) |
|---|---|---|---|---|
| GENDER | Male | - | 75.2 | 117 |
| | Female | - | 76.4 | 55 |
| AGE | AGE_ONE | 18-24 years | 75.0 | 36 |
| | AGE_TWO | 25-29 years | 82.8 | 64 |
| | AGE_THREE | 30-39 years | 68.4 | 57 |
| | AGE_FOUR | 40-65 years | 73.3 | 15 |
| JOB | JOB_1 | Employed | 71.7 | 53 |
| | JOB_2 | Student | 77.3 | 119 |
| EDUCATION | EDU_ONE | college/university | 71.9 | 32 |
| | EDU_TWO | Bachelors | 78.6 | 28 |
| | EDU_THREE | Masters | 80.0 | 55 |
| | EDU_FOUR | PhD | 71.9 | 57 |
| LICENSE | DRIVE_ONE | G1 | 86.7 | 15 |
| | DRIVE_TWO | G2 | 73.1 | 26 |
| | DRIVE_THREE | G | 69.0 | 87 |
| | DRIVE_FOUR | Other than Ontario | 86.4 | 44 |
| DRIVING EXPERIENCE | DRIVE_EXP_ONE | <2 years | 77.4 | 31 |
| | DRIVE_EXP_TWO | 2-5 years | 73.1 | 26 |
| | DRIVE_EXP_THREE | 5-10 years | 79.6 | 49 |
| | DRIVE_EXP_FOUR | >10 years | 72.7 | 66 |
| exp_give_before | NO | giveAway experience before | 72.1 | 86 |
| | YES | | 79.1 | 86 |
| famAV | NO | Familiarity about AVs | 75.8 | 33 |
| | YES | | 75.5 | 139 |
| MULTI_TASKING[1] | NO | visual manual mounted secondary tasks | 68.4 | 79 |
| | YES | | 81.7 | 93 |
| WEATHER | CLEARSKY/SUNNY | - | 78.6 | 84 |
| | RAINY | - | 72.7 | 88 |
| LIGHTING[1] | DAY | - | 73.3 | 86 |
| | NIGHT | - | 77.9 | 86 |
| TRAFFIC[1] | HEAVY_CONGESTION | - | 73.9 | 88 |
| | LIGHT_CONGESTION | - | 77.4 | 84 |

[1] Controlled variables; giveAway(%) shows percent of observations in which driving control is given to automation

The descriptive statistics provided in Table II offer valuable insights into the "giveAway" responses at the interval level. For example, observations from female participants accounted for 32 percent of the dataset (55 out of 172). Their "giveAway" response (42 out of 55, 76.4 percent) was slightly higher than that of male participants. Participants over the age of 40 constituted less than 9 percent of the overall population and exhibited the second-lowest "giveAway". Nearly 70 percent of the participants were students, demonstrating a higher inclination to give away control. Furthermore, over 65 percent of the participants held higher levels of education (EDU_THREE plus EDU_FOUR), with those having a master's degree displaying the greatest "giveAway". Likewise, drivers with licenses from jurisdictions other than the province of Ontario showed higher "giveAway" responses. Conversely, drivers with more than ten years of experience exhibited the lowest rate of giving away control. Previous experience with "giveAway" situations and familiarity with autonomous vehicles (AVs) also influenced participants' control-sharing choices. The multitasking scenario presented by visual-manual mounted secondary tasks, represented by reading a virtual newspaper or playing a maze game on a virtual mobile phone, particularly encouraged drivers to share control, as evidenced by its 82 percent "giveAway" response. Clear weather conditions resulted in a higher frequency of control sharing than rainy weather. During daytime driving, there was a 27 percent "no_giveAway" response, whereas nighttime driving had a 22 percent "no_giveAway" response, indicating that drivers were more likely to relinquish control during nighttime hours. Additionally, light traffic congestion corresponded to a relatively higher rate of "giveAway" than heavy traffic congestion.

## IV. MODELING APPROACH

This section provides an overview of the primary steps taken in the modelling approach to address the underlying research questions. Firstly, we discuss how we examine the individual and subpopulation-level unobserved variations in the binary choice of giving away control. Secondly, we discuss the process of obtaining the ordinal responses and constructing an ordinal logit model by incorporating the kinematic behaviour.

### A. Modelling Unobserved Heterogeneity

The binary dependent variable reflects the choice made by drivers to voluntarily hand over driving control to automation or, to put it another way, to switch to autopilot mode while driving. For analyzing unobserved heterogeneity in the binary outcomes, the following four steps are taken, and the listed statistical models are estimated:

*Step 1.* The binary logit model is implemented as a base model that assumes a linear relationship between the explanatory variables and the log odds of the binary outcome. All explanatory variables were dummy coded except the perceptual variable of locus of control (scores range from 0 to 13). Only statistically significant indicators are primarily used in the model's utility functions. The estimated parameters include the interaction effect through the compound variables besides the individual effects of each predictor variable (i.e., main effect). The primary limitation lies in its reliance on the assumption of independence from irrelevant alternatives (IIA), which has spurred researchers to explore advanced formulations.

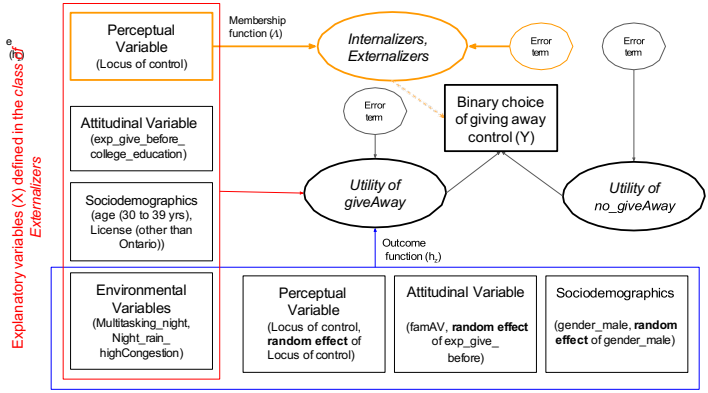

Fig. 2: Schematic Diagram of Latent Class Mixed Logit (LCML) Model

*Step 2.* The mixed logit (MIXL) model is estimated using a numerical integration technique to account for preference heterogeneity among the individuals in the sample. The random parameter draws were assumed to follow a normal distribution. MIXL helped to identify the covariates of interest in describing classes or to parameterize the membership function in the LCM model.

*Step 3.* The latent class model (LCM) is estimated to address the latent unobserved heterogeneity, assuming the population is divided into several segments, each with a distinct preference structure. It resembles MIXL, incorporates semiparametric specifications, and liberates the modeller from making distributional assumptions regarding individual heterogeneity. The path diagram of the latent class model is shown in Figure 2. This is a sort of saturated structure in which the membership variable $W$ is also directly influencing the outcome variable $Y$. Through LCM, unobserved heterogeneity is investigated in explanatory variables, parameters (preferences for factors associated with outcomes), and attributes (different model specifications for each class). The functional form is homogeneous for both classes (as $Z = 2$). Each segment follows the same data generation process, which follows binary logit form. A general form of the latent class model is as follows;

$$f(y) = \sum_{z=1}^{Z} f(y|z, X; \beta_z) P(z|W; \alpha) \quad (1)$$

Where:

$f(y)$ is the density function of an outcome variable $Y$ (dependent variable)
$z$ is a discrete subgroup indicator
$X$ is a vector of explanatory variables explaining $Y$
$W$ is a vector of membership variable explaining $z$
$\beta_z$ is a parameter vector for the outcome model
$\alpha$ is a membership parameter vector

$P(.)$ denotes a segment membership probability and $f_z(.)$ denotes the probability density function of the outcome $Y$ for segment $z$. The probability that individual $i$ with profile $X_i$ belongs to subgroup $z$ equals;

$$P_{i,z} = \frac{\exp(X_i \beta_z)}{\sum_{z=1}^{Z} \exp(X_i \beta_z)} \quad (2)$$

Let the behavioural model is a binary choice logit with alternative $j$ and choice situation $t$; then the probabilities can be expressed as.

$$Prob(y_{it,j} = 1|X_{it}, class = z) = \frac{1}{1 + \sum_{z=1}^{Z} \exp[X_{it,j} \beta_z]} \quad (3)$$

*Step 4.* Mixing in the latent segments is evaluated to address the response variations in the within-class. It possesses the virtue of combining the dual merits of both MIXL and LCM. The mixing in the latent class model can be introduced by the vector of random parameters ($\sigma_{ij}$) by the assumption that the ($\sigma_{ij}-s$) follow a normal distribution.

$$Prob(y_{it,j} = 1|X_it, class = z) = \frac{1}{1 + \sum_{z=1}^{Z} \exp[X_{it,j}(\beta_z + \sigma_{ij})]} \quad (4)$$

### B. Ordinal Responses in Automated Driving Adoption

The automated driving adoption is evaluated based on the duration and proportion of automated driving in the L3 vehicle trip in a mixed-traffic environment. The analysis focuses explicitly on trips where drivers transferred control to automated driving only once. Trips involving repeated transitions and multiple instances of control switching, encompassing the behaviour of taking back control from automation, are not considered. Furthermore, the small proportion, which accounted for less than 20 percent of the total observations, and its different impact on the analysis motivated the decision to exclude trips with multiple control transitions.

In the first step, Linear regression is estimated as a base model to measure the effect of kinematic parameters, attitudinal variables, sociodemographics and scenario-related control variables on the automated driving proportion. Second, the Jenks Natural Breaks (JNB) classification method [26] is applied to categorize the automated driving proportion to ensure a more interpretable and quantifiable analysis. JNB stands out from other commonly used clustering methods due to its notable robustness against outliers and its ability to effectively distribute one-dimensional data into natural breaks. The discretization of data aims to minimize the variance within each group while maximizing the variance between different groups. JNB has categorized the automated proportion into three levels of adoption: low category for automated driving proportions up to 39 percent, medium level for proportions between 40 and 66 percent, and high level for proportions between 67 and 95 percent. These categories have been created with approximately equal numbers of observations. Finally, the ordinal logit model is estimated to measure the category-specific impacts of explanatory variables on automated driving adoption.

The utility ($U_n$) of unobserved ordinal responses ($r_k$) remains constant between ordinal categories, leading to the parallel regression assumption. A set of thresholds ($\delta_k$) associating the latent variable ($U_n^*$) to an ordinal response is estimated in addition to regression coefficients ($\beta-s$) of explanatory variables ($x_n$). The estimation of optimal parameters and the development of a predictive ordered logit model from the data are accomplished using the Maximum Likelihood method. The choice probability of selecting the $k^{th}$ category from an ordinal choice set is determined by the difference of cumulative probabilities, as follows:

$$P(U_n = r_k) = P(U_n^* > \delta_k) - P(U_n^* > \delta_{k-1})$$
$$= \left(\frac{1}{exp(\beta x_n - \delta_k)}\right) - \left(\frac{1}{exp(\beta x_n - \delta_{k-1})}\right) \quad (5)$$

Both the linear regression and ordinal logit models employ identical explanatory variables. These variables consist of manual driving acceleration as a kinematic parameter, familiarity with autonomous vehicles (AVs) as an attitudinal variable, sociodemographics, and scenario-related variables. While linear regression offers valuable insights into the direction, magnitude, and significance of relationships, assuming linearity, the ordinal logit model explores the order, magnitude, and odds ratios of explanatory variables within the ordinal outcome framework.

## V. IMPLEMENTATION DETAILS

The implementation begins by using correlation, causation, and basic statistical analysis to validate the study's initial intuitions and prior hypotheses. For example, by conducting multivariate correlation analysis, we obtained insights into the direction and magnitude of the relationships between independent and dependent variables, as well as among the independent variables themselves. This initial analysis helped us identify the significant exploratory variables in a preliminary manner. Once we understood the initial intuitions and hypotheses through these analytical methods, we opted to start with simpler discrete choice models and specifications. This approach allowed us to obtain good initial values and reduce the estimation cost of more complex models. Ultimately, we developed the final models to investigate the underlying factors and their impact on an individual's WTG.

To implement the discrete choice models incorporating the selected variables, a Python library for discrete choice modelling called 'Biogeme' was utilized. The systematic component of the utility function in underlying models was established by including perceptual and attitudinal variables, sociodemographic factors, non-behavioural variables such as weather, lighting, multi-tasking, and traffic congestion, as well as a kinematic parameter. The dependent variables involved in the present analysis are: Y1: 'Yes'(1) if the participant opts to give control to automated driving, otherwise 'No'(0). Y2 (a): the proportion of time (sec) participants spend using automated driving within an L3 vehicle trip (continuous variable). Y2 (b): low, medium, and high category of automated driving adoptions based on JNB classification ('1' for low, '2' for medium, '3' for high).

## VI. RESULT AND ANALYSIS

This section presents the outcomes and analysis derived from the underlying models. The analysis of the binary choice potentially reflects the user acceptance and trust in automated driving, while the examination of the proportion and level of automated driving adoption allows for an estimation of how various driving settings may influence adoption patterns.

### A. Binary Choice of Give Away Control

Table III presents the parameter estimation results of binary logit (BL), mixed logit (MIXL), and latent class mixed logit (LCML) models. The estimated coefficients of explanatory variables exhibit two response types: 'likely to give' for positive estimates and 'likely to not give' control for negative estimates. For example, the positive and significant estimates of ASC_Give indicate that the alternative of giving away control has a higher level of utility than the reference alternative (i.e., not giving control). It means that, with all else equal, drivers were primarily more willing to relinquish their driving control to automation. This behaviour aligns with the findings that individuals regard great value in the benefits of automated driving adoption on highways [27]. Furthermore, the estimation of explanatory variables considers fixed (mean) effects, mixing effects, and compound effects. In the BL model, the emphasis is on fixed and compound effects, whereas the MIXL model incorporates the mixing effect by estimating the parameters of standard deviation variables to assess the level of indifference towards a choice. The presence of significant taste heterogeneity, if revealed by the MIXL model, indicates that the preference for the giveaway option is not uniform across individuals.

The locus of control index (LCI) estimation in the BL and MIXL models negatively affects the driver's willingness to give away control. Therefore, drivers with Higher LCI (external locus) are less likely to use the automated mode. However, the mixing effect of LCI_S verdicts considerable variations (±0.15 from the mean estimate of -0.25), which means that the behaviour of LCI is highlighting the taste variation among drivers. For instance, the confidence interval here is (-0.4, -0.1), indicating considerable taste variations with the external locus variable. This implies that we cannot overlook the response variations and assume that individuals with an external locus invariably align with the hypothesized trust and adoption of automated technology, as proposed in the literature. Similar to this, the effect of the male_with_Glicense_S parameter (male drivers with a G driving licence) varies among drivers. Additionally, drivers with an age limit of 30 to 39 years are less likely to give control to automation. Drivers do not feel comfortable being automated when it is rainy at night, and there is heavy traffic (night_rain_highcongestion).

The estimate of compound indicator exp_give_before_college_education reflects that the drivers who have already gained control-giving experience and whose highest level of education is a college diploma are more likely to use the automated mode. The magnitude of the estimated coefficient in the BL model is 1.92, indicating the strength of the effect of the compound variable on the likelihood of the binary outcome. Specifically, its odds ratio (i.e., exp(coef(results)))

TABLE III: Estimation results of the behavioural choice models for giving away control

| Parameters | Binary Logit (BL) | | Mixed Logit (MIXL) | | Latent Class Mixed Logit (LCML) | | | |
|---|---|---|---|---|---|---|---|---|
| | Estimate | Rob. t-stats | Estimate | Rob. t-stats | Estimate | Rob. t-stats | Estimate | Rob. t-stats |
| | | | | | Externalizers | | Internalizers | |
| Alternative Specific Constant (ASC_Give) | 2.33 | 3.47 | 2.96 | 3.61 | 0.79 | 2.85 | 8.31 | 8.32 |
| **Perceptual Variable** | | | | | | | | |
| locus of control index (LCI) | -0.20 | -2.20 | -0.25 | -2.43 | - | - | 9.04 | 14.80 |
| **Attitudinal Variables** | | | | | | | | |
| familiarity about AVs (famAV) | - | - | - | - | - | - | 16.6 | 13.5 |
| exp_give_before_college_education[a] | 1.92 | 2.14 | 1.88 | 1.82* | 2.15 | 2.46 | - | - |
| **Sociodemographics** | | | | | | | | |
| age (30-39 yrs) | -0.87 | -1.90* | -1.17 | -2.13 | -1.06 | -2.24 | - | - |
| license (other than Ontario) | - | - | - | - | 1.40 | 2.30 | - | - |
| gender_male | - | - | - | - | - | - | -15.50 | -11.1 |
| male_with_Glicense[a] | -0.85 | -2.08 | -0.77 | -1.67* | - | - | - | - |
| **Environmental Variables** | | | | | | | | |
| multitasking_night[a] | 2.31 | 3.60 | 2.43 | 2.78 | 2.31 | 3.17 | 16.6 | 13.5 |
| night_rain_highcongestion[a] | -1.04 | -1.77* | -1.35 | -1.65* | -1.61 | -2.20 | 15.9 | 11.2 |
| < 5yrDexp_day_sun[a] | 1.27 | 1.95* | 1.39 | 1.67 | - | - | - | - |
| **Attributes (Standard deviation for Taste Heterogeneity)** | | | | | | | | |
| Std dev of locus of control index (LCI_S) | - | - | 0.15 | 1.52** | - | - | -0.045 | -18.2 |
| Std dev of males with Glicense (male_with_Glicense_S) | - | - | -1.62 | -1.91* | - | - | - | - |
| Std dev of males (gender_male_S) | - | - | - | - | - | - | -0.05 | -5.03 |
| Std dev of experience of give away before (exp_give_before_S) | - | - | - | - | - | - | 0.013 | 1.25** |
| **Class Membership Function** | | | | | | | | |
| coef_intercept | - | - | - | - | -27.70 | -5.37 | - | - |
| coef_Locus | - | - | - | - | 9.77 | 3.97 | - | - |
| **Performance Indicators:** | | | | | | | | |
| Number of parameters | 8 | | 10 | | 17 | | | |
| Akaike Information Criterion | 172.560 | | 175.776 | | 181.458 | | | |
| Bayesian Information Criterion | 197.741 | | 197.970 | | 234.965 | | | |
| Rho-square-bar | 0.102 | | 0.127 | | 0.239 | | | |

[a] Compound Variables, * Not statistically significant at 95% confidence level, ** Not statistically significant at 90% confidence level

= $e^{1.92}$) shows that there is a 6.92-fold increase in the odds of choosing the automated mode for every one-unit increase in this compound variable, compared to the reference level (i.e., no control-giving). In other words, individuals who have higher levels of "exp_give_before_college_education" are approximately 6.92 times more likely to give the control compared to those with lower levels, all else being equal. Here, a t-value of 2.14 suggests that the estimated coefficient is 2.14 standard errors away from zero and is statistically significant at a 95 percent confidence level. Moreover, another compound variable, multitasking_night, exhibits a positive relationship with the binary outcome but with a higher magnitude and statistical significance. This suggests drivers are likelier to opt for the automated mode in the nighttime multitasking scenario. One possible explanation is that drivers choose to 'giveAway' control of the vehicle when greater concentration/attentiveness is required or when they want to perform non-driving-related tasks (NDRTs). The latter hypothesis is supported by the significant occurrence of drivers engaging in smartphone usage for NDRTs during L3 automated driving, as observed in [28].

In the LCML, latent segments were developed by the latent class model based on the taste heterogeneity variable identified from the MIXL model. The underlying philosophy of this approach lies in acknowledging and accounting for taste heterogeneity, which allows for a deeper understanding of the diverse preferences within a population. It recognizes that individuals possess distinct preferences and that assuming a homogeneous population can result in oversimplified and less precise findings. The latent class membership function is defined by locus indicator (coef_Locus) within the first class that divides the population into two latent classes. It assumes that two distinct groups capture the choice of giving away control. The dichotomized typology of the Rotter locus is highly applicable in research [29]. It provides a binary framework to explore and compare the preferences between these two groups. The group of drivers within the first class are referred to as Externalizers based on the higher LCI and externality locus (significant estimate of coef_Locus is 9.77). On the contrary, the intercept of Locus (coef_intercept) being -27.7 indicates that individuals with a lower locus of control index are more likely to belong to the class of internality locus, which we referred to as Internalizers. ASC_Give of these classes indicates that more participants belong to the group of Internalizers. The probability of an individual being in first-class (Externalizers) increases as the locus index (LCI) increases. Given significant and positive ASC_Give for both classes, with all else equal, participants are more likely to give control to automation.

The systematic components differ between classes indicating that the explanatory variables have different effects and significance in shaping choices within each class. As for the first class, being in the age group of 30 to 39 years or driving in the environment of night_rain_highcongestion tends to impact their willingness to give control negatively. On the other hand, drivers without Ontario licensing among Externalizers are more willing to give control. Additionally, multitasking during nighttime has the highest influence on the choice of Externalizers whether they give control. For the internalizers, an attitudinal variable of famAV is highly significant in their willingness to give control. Drivers already familiar with automated driving are more likely to give control. Similar to externalizers, internality locus individuals highly tend to give control in the scenarios when they are allowed to multitask at nighttime. Additionally, higher LCI of internalizers positively affects them to relinquish driving control. However, males among internalizers were less likely to give control.

The variability of the LCI_S parameter is found to be significant within the internalizers class (estimate(t-value) is -0.045(-18.2)), indicating its non-homogeneous impact on individual choices. However, when considering the entire population in the MIXL, the LCI_S parameter (estimate(t-value) is 0.15(1.52)) does not exhibit significant heterogeneity. The differences in significance, signs, and weights of the random effects between the two models highlight the importance of considering latent classes in capturing heterogeneity and understanding how the effects of parameters vary across different segments of the population. Moreover, the scale and random parameters for the externalizer class are absent, indicating that they do not significantly contribute to the heterogeneity observed in the data. One possible behavioural interpretation that internalizers demonstrate more heterogeneity compared to externalizers could be because of their personal attribution of outcomes to their abilities and choices, leading to diverse perspectives, adaptive strategies, and varied decision-making approaches. Furthermore, although the internalizers show statistical significance in three random attributes, their estimated values are relatively small. These findings suggest that the existing two latent classes effectively capture the heterogeneity in the data, and the level of unobserved heterogeneity within each latent class is minimal.

In terms of the model fit, BL and MIXL models had more or less the same goodness of fit indices, i.e. rho-square bar, AIC, and BIC, showing similar performance. LCML model outperforms the binary and mixed logit models regarding goodness of fit. It reflects that the LCML model can more accurately assign observations to their respective latent classes. Furthermore, the LCML stands out as the most complex among the three models, as evidenced by its higher number of parameters and AIC/BIC values. This highlights the trade-off that exists between the complexity of a model and its goodness of fit.

### B. Levels of Automated Driving Adoption

The analysis of the level of automated driving adoption is required for a broader understanding of an individual's WTG regarding adoption patterns across different driving settings. The estimation outcomes for the ordered logit (OL) model and linear regression (LR) are shown in Table IV. The models have estimated the effect of kinematic parameters, attitudinal variables, sociodemographics, and scenario-related control variables on the proportion and levels of automated driving adoption. In the LR model, the dependent variable (DV) was the automated driving proportion in L3 vehicle trips which was continuous and normalized (as defined by Y2 (a) in the previous section). In contrast, the DV for the OL model was based on the JNB-classified categories of low, medium, and high levels of automated driving proportion (as defined by Y2 (b) in the previous section). Regarding the parameters of explanatory variables, positive coefficients show that the associated variable increases the likelihood of a higher level of automated driving adoption, while negative coefficients show the opposite.

In the LR model, the negative intercept/constant (-0.66) represents the proportion of time spent in automated mode is expected to be lower than the manual driving proportion when all predictor variables are set to zero. Given that acceleration during manual driving positively impacts the dependent variable, the greater change in speed was likely to lengthen the automated driving duration. Two interpretations could be derived from changes in velocity linked to the adoption of automated driving, either due to chaotic control switching decisions or due to nighttime and multitasking, as both environmental conditions positively affect automated driving adoption. Similarly, those familiar with AVs (famAV), drivers with less than two years of experience behind the wheel, and those over 40 were also more inclined to highly share the control with automation. Compared to men, women were more likely to employ the automated mode. Last but not least, the automated mode appears to be used less by drivers who do not possess an Ontario driver's licence.

TABLE IV: Models for automated driving proportion

| Parameters | Linear Regression | | Ordinal Logit | |
| --- | --- | --- | --- | --- |
| | Estimate | Rob. t-stats | Estimate | Rob. t-stats |
| constant | -0.66 | -3.11 | - | - |
| **Kinematic parameter** | | | | |
| manual driving acceleration | 0.48 | 6.73 | 1.80 | 2.98 |
| **Attitudinal variable** | | | | |
| famAV (familiarity about AVs) | 0.40 | 2.06 | 1.17 | 1.86 |
| **Sociodemographics** | | | | |
| < 2years_DrivingExperience | 0.33 | 1.51 ** | 0.94 | 1.69 * |
| age (40-60 yrs) | 0.89 | 3.44 | 3.51 | 3.73 |
| gender_female | 0.25 | 1.67 * | 0.42 | 0.84 |
| license (other than Ontario) | -0.59 | -3.60 | -1.58 | -2.77 |
| **Scenario related variables** | | | | |
| night | 0.28 | 2.05 | 0.89 | 1.93 * |
| multitasking | 0.24 | 1.74 * | 0.48 | 1.10** |
| **Threshold and differences** | | | | |
| tau1 | - | - | 0.63 | 1.02** |
| delta2 | - | - | 2.25 | 5.59 |
| delta3 | - | - | 17.20 | 8.34 |
| **Performance Indicators** | | | | |
| Number of Parameters | 9 | | 11 | |
| R-squared | 0.56 | | 0.42 | |
| Adj. R-squared | 0.52 | | 0.34 | |
| Log-Likelihood | -106.87 | | 110.41 | |
| Akaike Information Criterion | 231.70 | | 175.77 | |
| Bayesian Information Criterion | 255.70 | | 205.07 | |

\* Not statistically significant at 95% confidence level
\*\* Not statistically significant at 90% confidence level

In the OL model, the three-ordered dependent variable is influenced by the same explanatory variables in a similar pattern but with different magnitudes. We gain an understanding of category-specific impacts through the OL model. The category/level of automated driving adoption can be predicted based on the threshold parameters and the utility function. For instance, the threshold between the first and second levels is represented by tau1. A higher tau1 value (+ 0.63) indicates that the threshold for transitioning from the first level to the second level is located further along the latent scale. This suggests that higher values on the latent scale are needed for an individual to have a higher probability of belonging to the second level than the first. Additionally, the significant delta values indicate a substantial and significant distance between adjacent categories. In the underlying model, the distance between the thresholds for the second and first levels is represented by delta2, whereas the distance between the thresholds for the third and second levels is represented by delta3. In summary, the utility function or latent variable is projected to have a medium level of automated driving adoption if it is more than tau1 and less than tau2 (tau1+ delta2).

According to the estimates obtained from the OL model, individuals aged 40 and above tend to belong to the highest level of automated driving adoption. On the other hand, individuals without Ontario licensing are more likely to be categorized in the lowest level of automated driving adoption. It might be because immigrants exhibit more carefulness in various aspects of life. They may approach the use of automated technology with more caution. Moreover, factors such as manual driving acceleration and familiarity with AVs have a significant positive effect on the adoption of automated driving. Females exhibit a higher tendency to embrace automated driving compared to males. These findings align with those observed in [30], indicating that women tend to favour a medium level of automation. Additionally, consistent with our previous findings in binary logit analysis, engaging in multitasking and driving in nighttime environments are encouraging factors for adopting automated driving.

The observable explanatory variables were the same in both the LR and OL models. The adjusted R-squared value of 0.52 in the LR model shows that the predictor variables account for the dependent variable's variability to a degree of about 52 percent in LR. However, the lower AIC and BIC values for the OL model point to a more efficient model with a better model fit. This suggests that the OL model captures the underlying relationships between the predictors and the ordinal DV more effectively, given the nature of the data.

## VII. Concluding Discussion

This study delved into the objective analysis of heterogeneity in willingness to give control (WTG) to automated driving, focusing on SAE Level 3 vehicles. We used a virtual and immersive environment with a driving rig to execute the exploratory simulator study in different driving settings. By employing latent class mixed logit and ordinal logit models on the collected data, the study investigated the presence of unobserved heterogeneity within and across classes and estimated the responses towards unobserved levels of automated driving adoption. The obtained results highlighted the significant factors influencing the choice and levels of automated driving adoption. We found that drivers choose to 'giveAway' control of the vehicle when greater concentration/attentiveness is required (e.g., in the nighttime) or when they are interested in performing non-driving-related tasks (NDRTs). From the analysis of Rotter's Locus of control, we came up with the dichotomized typology in which the class of internalizers had a non-homogeneous impact on an individual's choices, and it is variably aligned with the hypothesized trust and adoption of an automated technology. Our findings contribute to better identifying sub-populations with varying preferences of WTG toward conditional automated driving.

The underlying two analyses are interrelated, providing complementary information. The binary choice analysis provides insights into individual-level preferences, while the examination of adoption patterns offers a broader perspective on the factors influencing the category-specific adoption of automated driving. Integrating the insights from both analyses allows researchers and policymakers to gain a comprehensive understanding of an individual's WTG, including the identification of factors that influence users' decision-making regarding automated driving. Moreover, this integrated approach facilitates a deeper comprehension of the overall adoption patterns across various driving contexts. By leveraging these combined findings, policymakers and auto OEMs can improve driver monitoring systems and develop effective control transition strategies to promote the widespread usage of automated driving. The findings from the present behavioural analysis can potentially inform insurance policy design by facilitating the development of personalized insurance policies. The valuable information gleaned from the study can assist in developing a business-to-business (B2B) insurance solution tailored explicitly to automakers, as they are held liable and bear responsibility for the performance of the automated driving system.

The present study incorporates simultaneous analysis to expand the in-depth analysis by estimating the choice model parameters and exploring the levels of willingness to give control with an automated system. The approach of simultaneously estimating the multiple-choice models has been utilized in previous studies for two primary purposes. The objective either encompassed identifying the preferred model or integrating outcomes from hybrid models to enhance predictive capabilities and improve overall model performance. However, our study makes an additional contribution by utilizing insights from one model to develop an advanced model. For instance, we employed MIXL to identify the membership variable, which assisted in developing the latent classes in the LCML model.

The main limitation of this research is considering the particular trips in the analysis of automated driving proportion where drivers have just once handed over the control to automated driving. In future studies, we aim to model the automated driving adoption in trips where the driving control is repeatedly switched. Moreover, the different utility equations of *Internalizer* and *Externalizer* classes limit the possibility of direct comparisons between them. Future studies aim to explore consistent indicators to facilitate meaningful comparisons between the classes. While the underlying analysis has extensively examined significant explanatory variables, there remain potential areas for further research that can merit exploration. One such area is the analysis of surrogate safety measures, such as Time-to-collision (TTC) and lane change duration. The analysis of these measures is crucial for assessing safety implications and gaining valuable insights into how individuals adapt to and respond to automated systems. In future work, we intend to enhance the behavioural analysis of L3 control transitions by increasing the number of observations. We also aim to investigate the mixing effect in ordinal logit for the best estimation and forecasting. Furthermore, in conclusion, we recommend that researchers carefully consider the trade-off between model complexity and data fit when interpreting and selecting the most suitable model.

## BIBLIOGRAPHY

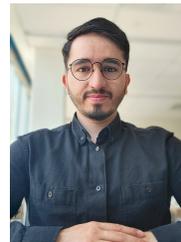

**Muhammad Sajjad Ansar** is currently a Ph.D. candidate at the LiTrans Lab of Toronto Metropolitan University. He holds a MASc degree (2020) in Transportation Engineering from Southeast University, China. He obtained a fully funded master's degree scholar award from the Chinese Scholarship Council (CSC). He received B.Eng. in 2018 from the University of Engineering and Technology, Lahore, Pakistan. His current work focuses on behavioural modelling, connected and automated vehicles, and traffic safety analysis.

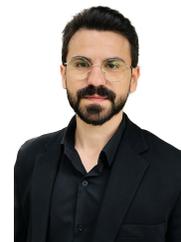

**Nael Alsaleh** is currently a postdoctoral researcher at the Mobility Network, University of Toronto. Nael completed his B.Sc. in Civil Engineering in 2015 and M.Sc. in Transportation Engineering in 2017, from Jordan University of Science and Technology (JUST), Jordan. His M.Sc. thesis won the first-place award in The Best Master's Thesis Competition at JUST for the 2017-2018 academic year. Alsaleh earned his Ph.D. in Transportation Engineering in 2022 from Toronto Metropolitan University, Canada. His research interests include travel demand modelling, connected and autonomous vehicles, on-demand shared mobility services, and transportation systems simulation.

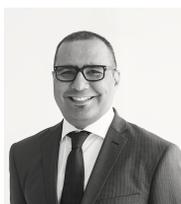

**Bilal Farooq** received B.Eng. degree (2001) from the University of Engineering and Technology, Pakistan, M.Sc. degree (2004) in Computer Science from Lahore University of Management Sciences, Pakistan, and Ph.D. degree (2011) from the University of Toronto, Canada. He is the Canada Research Chair in Disruptive Transportation Technologies and Services and an Associate Professor at Toronto Metropolitan University. He received Early Researcher Award in Quebec (2014) and Ontario (2018), Canada. His current work focuses on the network and behavioural implications of emerging transportation technologies and services.